\documentclass[a4paper,11pt]{article}

\usepackage[utf8]{inputenc}
\usepackage[T1,T2A]{fontenc}
\usepackage{amsmath,amsfonts,amssymb}

\usepackage{cite}

\usepackage{graphicx}
\usepackage{wrapfig}
\usepackage{hyperref}

\usepackage[bottom]{footmisc}

\usepackage{geometry}
\newcommand{\lb}[0]{\left(}
\newcommand{\rb}[0]{\right)}
\newcommand{\lsb}{\left[}
\newcommand{\rsb}{\right]}
\newcommand{\qrq}{\quad\Rightarrow\quad}
\newcommand{\pz}{\partial_z}
\newcommand{\veps}{\varepsilon}

\begin{document}

\renewcommand*{\thefootnote}{\fnsymbol{footnote}}

\begin{center}
{\large\bf  Gluon string breaking and meson spectrum in the holographic Soft Wall model}
\end{center}
\bigskip
\begin{center}
{ Sergey Afonin\footnote{E-mail: \texttt{s.afonin@spbu.ru}.}
and Timofey Solomko}
\end{center}

\renewcommand*{\thefootnote}{\arabic{footnote}}
\setcounter{footnote}{0}

\begin{center}
{\small\it Saint Petersburg State University, 7/9 Universitetskaya nab.,
St.Petersburg, 199034, Russia}
\end{center}

\bigskip

\begin{abstract}
We propose a general method for finding a string-like meson spectrum which is based on
a certain condition for the breaking of closed gluon string. A model is constructed that
successfully realizes the proposed method.
The model is based on the holographic Soft Wall model for QCD and the use of the Wilson confinement criterion.
We applied our approach to the vector and scalar cases and obtained numerical predictions for the intercepts
of corresponding Regge like radial meson spectra. A good agreement is obtained both with the existing
experimental data and with some other known phenomenological approaches.
Remarkably, our closed string breaking condition has two branches. We argue that they should correspond
to states of opposite parity. The Wilson confinement criterion leads then to a natural mass splitting
between parity partners. The constructed model represents thus the first example of bottom-up holographic model
in which the effects of chiral symmetry breaking can emerge automatically, i.e., without additional assumptions
taken outside of the holographic approach.
\end{abstract}

\bigskip


\section{Introduction and general idea}

Some time ago, Andreev and Zakharov made an interesting observation~\cite{Andreev:2006ct}
that the Cornell heavy-quark potential,
\begin{equation}
\label{cornell}
V(r)=-\frac{\kappa}{r}+\sigma r + C,
\end{equation}
can be reproduced within the so-called Soft-Wall (SW) holographic model~\cite{son2,andreev}.
Later the SW model became a popular AdS/QCD approach to modeling the non-perturbative strong interactions ---
the hadron Regge spectroscopy, hadron form-factors, distribution amplitudes, QCD thermodynamics,
etc. (many modern references are given in the recent paper~\cite{Afonin:2021cwo}). In the present Letter,
we extend the approach of Ref.~\cite{Andreev:2006ct} and use it to construct a model for calculation of the
intercept of linear radial spectrum of light mesons.

The linear Regge trajectories arise in the string approach to QCD. The string picture of QCD, in its turn,
looks most naturally in the limit of large number of colors (called also planar),  \(N_c\to\infty\), of QCD~\cite{hoof,wit}.
QCD simplifies considerably in the planar limit, in particular, the spectrum consists of infinite number of narrow
(the decay width is suppressed by a $\mathcal{O}(N_c^{-1})$ factor)
colorless glueballs and mesons with $\mathcal{O}(1)$ masses.
The appearance of holographic QCD approach, which is inherently a large-$N_c$ approach, has awakened interest in
finding a string description of strong interactions.

If the string picture is valid, the spectrum should be string-like, at least in the first approximation.
Let the radial spectrum of glueballs be
\begin{equation}
\label{spectrum1}
  M^2_\text{gl}(n)=a(n+1),\quad n=0,1,2,\dots.
\end{equation}
A concrete value of intercept is not relevant and we consider the simplest one.
The spectrum of mesons is expected to be different,
\begin{equation}
\label{spectrum2}
  M^2_\text{mes}(n)=a'(n+1+b),
\end{equation}
i.e., with a different slope and intercept. The latter depends on the quantum numbers of quark-antiquark pair.
Note in passing that the spectrum~\eqref{spectrum2} qualitatively arises in the planar two-dimensional QCD~\cite{hoof2D}
and also it seems to hold semiquantitatively in real light mesons with universal slope~\cite{Anisovich:2000kxa,Bugg:2004xu,Klempt:2007cp,
Li:2004gu,Shifman:2007xn,Afonin:2006vi,Afonin:2006wt,Afonin:2007jd,Afonin:2007aa}, with the
intercept dependent on spin-parity in a certain way~\cite{Afonin:2006vi,Afonin:2006wt,
Afonin:2007jd,Afonin:2007aa}.

When the quarks are in the fundamental representation of $SU(N_c)$ gauge group, the gluons
absolutely dominate in the strong dynamics, thus the quark effects are $\mathcal{O}(1/N_c)$ suppressed and can be completely neglected
in the first approximation. Imagine that we know the glueball spectrum~\eqref{spectrum1} in the leading
order in \(N_c\to\infty\). We will consider the following problem: How can we get the spectrum~\eqref{spectrum2} from~\eqref{spectrum1}
in this leading order in $N_c$ where the quark dynamics is suppressed? This problem is partly relevant to the AdS/QCD models because the
holographic approach is formulated exclusively in the leading order in \(N_c\to\infty\).

\begin{figure}[b]
  \begin{center}
  \includegraphics[width=0.7\textwidth]{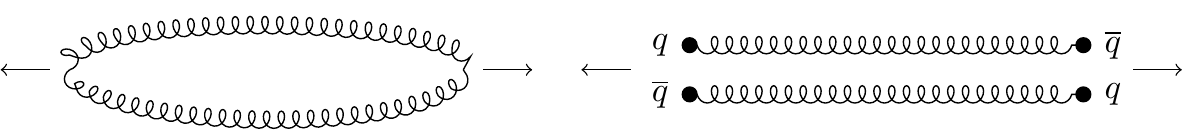}
  \end{center}
  \caption{A schematic view of a closed gluon string vs. two open strings.}
  \label{string_fig}
\end{figure}

Within the string picture of pure gluodynamics, the glueballs represent closed strings. If we stretch such a string,
its energy grows linearly,
\begin{equation}
\label{pot}
  E\simeq\sigma r + \text{const},
\end{equation}
where \(\sigma\) is the string tension. The introduction of quarks into consideration leads
to the possibility of breaking of closed strings to open strings. The stretching of a closed string costs then same
energy as the stretching of two open strings with quark-antiquark pair at the ends,
see Fig.~\ref{string_fig}. Thus, the tension of open string is expected to be twice
smaller than that of the closed string,
\begin{equation}
\label{rel}
  \sigma_\text{mes}=\frac{1}{2}\sigma_\text{gl}.
\end{equation}

The radial spectra~\eqref{spectrum1} and~\eqref{spectrum2} should emerge after quantization of the corresponding
strings. The slopes of those spectra are proportional to the string tension (this is evident from the mass dimension),
for instance, $a'=2\pi\sigma$ in the case of Nambu-Goto string.

Suppose that the quarks are introduced into gluodynamics leading thus to the possibility of open string creation.
How can one deduce the spectrum of quark-antiquark pairs from the spectrum~\eqref{spectrum1}?
The first guess follows from~\eqref{rel} and consists in the replacement $a\rightarrow a'=a/2$.
But this is not enough because the intercept is expected
to change as well, $n+1\rightarrow n+1+b$. Suppose further that we have a dynamical model for planar gluodynamics
which is able to reproduce the confinement behavior~\eqref{pot} at large distances and can incorporate
the dependence $\sigma(b)$ on the external parameter $b$ that we relate with quark effects.
The closed gluon string tension $\sigma(b)$ should be a decreasing function of $b$
because the fermions contribute to the strong dynamics with opposite sign.
And if at some point $b=b_0$ the tension $\sigma(b)$ is halved we get a tension that, according to~\eqref{rel},
is characteristic to the open string. Our basic idea is that the realization of the
corresponding condition,
\begin{equation}
\label{sigma_ratio}
  \sigma(b_0)=\frac{1}{2}\sigma(0),
\end{equation}
triggers the closed string breaking and formation of open string(s).

We thus propose the following answer to the question above: In order to find the string-like radial meson spectrum~\eqref{spectrum2}
from the glueball spectrum~\eqref{spectrum1} we should find the value of $b=b_0$ from the closed string breaking condition~\eqref{sigma_ratio}.
Since $a\sim\sigma$, the meson spectrum will be given by
\begin{equation}
\label{spectrum3}
  M^2_\text{mes}(n)=\frac12 a(n+1+b_0).
\end{equation}


Below we construct an explicit model realizing the proposed idea for prediction of
string-like meson spectrum in the large-$N_c$ limit and briefly compare our results with
the phenomenology and some other approaches.

The paper is organized as follows. The SW holographic model and its
generalization to the arbitrary intercept parameter is recalled in Section~2. In Section~3, we first give a brief review
on calculation of linearly rising potential within the SW model, then extend the corresponding technique to
the generalized SW model, and after that construct our model for finding the intercept in the case of vector mesons.
The results obtained are compared with the existing experimental data on light mesons and argued to be consistent both
with these data and with some known phenomenological approaches.
The scalar case is analyzed in Section~4. Section~5 is devoted to discussions and we conclude in Section~6.
Some technical details are given in Appendices.

\section{The SW holographic model}

Our starting point will be the well-known phenomenological SW holographic model.
The original SW model put forward by Son et al.~\cite{son2} is defined by the action
\begin{equation}
\label{SW}
  S=\int d^4xdz\sqrt{g}e^{-cz^2}\mathcal{L},
\end{equation}
where \(g_{MN}\) is the metric of the Poincare patch (\(z>0\)) of the AdS\(_5\) space
\begin{equation}
  ds^2=g^{MN}dx_Mdx_N=\frac{R^2}{z^2}\lb\eta^{\mu\nu}dx_\mu dx_\nu-dz^2\rb,\quad
  \eta_{\mu\nu}=\text{diag}\lbrace1,-1,-1,-1\rbrace,
\end{equation}
\(R\) is the radius of the AdS\(_5\) space and \(z\) is the fifth (called holographic) coordinate.
The function $e^{-cz^2}$ represents the dilaton background ($c$ is a constant) and
$\mathcal{L}$ is the Lagrangian density of some free field in AdS\(_5\) space which, by assumption, is
dual on the AdS\(_5\) boundary to some QCD operator. The model is defined in the probe
approximation, i.e., the metric is not backreacted by matter fields and dilaton (there exist also various dynamical
SW approaches, we will follow the standard probe approximation --- the backreaction
is assumed to be suppressed in the large-$N_c$ limit within which the holographic approach is formulated~\cite{Afonin:2021cwo}).
In the case of free vector fields, the mass spectrum of the model is given by~\cite{son2}
\begin{equation}
\label{spSW}
  M^2(n)=4|c|(n+1).
\end{equation}

There exists another formulation of SW model proposed almost simultaneously by Andreev~\cite{andreev}, in which the dilaton background
is replaced by a modified metric,
\begin{equation}
\label{az_metric}
  g_{MN}=\text{diag}\left\lbrace\frac{R^2}{z^2}h,\dots,\frac{R^2}{z^2}h\right\rbrace,\quad
  h=e^{-2cz^2}.
\end{equation}
The vector spectrum of this model coincides with~\eqref{spSW}.
The third formulation of SW model consists in incorporation of certain $z$-dependence into the 5D mass.
Namely, if the infrared modification of this mass has the form of (we set $R=1$)
\begin{equation}
\label{mz}
m_5^2(z)=m_5^2(0)+bz^2+c^2z^4,
\end{equation}
the spectrum will be linear~\cite{no-wall}, in particular it reduces to~\eqref{spSW} for $b=0$.
Various interrelations between different formulations of SW holographic model
were thoroughly studied in the recent work~\cite{Afonin:2021cwo}. Roughly speaking, these three formulations
are related by certain redefinitions of fields in the 5D action.
For instance, if a SW model is given by the action
\begin{equation}
\label{Bz}
  S=\int d^4x dz\sqrt{g}\,B(z)\,\mathcal{L},
\end{equation}
where $B(z)$ is some $z$-dependent dilaton background specifying the model,
this model can be reformulated as a SW model with rescaled fields
\begin{equation}
  S=\int d^4x dz\sqrt{\tilde{g}}\,\tilde{\mathcal{L}},
\end{equation}
and modified metric
\begin{equation}
\label{tr}
  \tilde{g}_{MN}=B^{(3/2-J)^{-1}}g_{MN}.
\end{equation}
Here $J$ denotes the value of spin of 5D field in the Lagrangian density $\mathcal{L}$.
It must be emphasized that all three formulations lead to identical classical equations of motion, hence,
to the same correlation functions and mass spectra. Since the SW holographic model remains purely phenomenological
construction, we still cannot judge {\it a priori} which formulation is more fundamental.
For our subsequent analysis, however, only the formulation with the modified metric is suitable.

One can construct a generalization of the standard SW model~\eqref{SW} which will be crucial for our further analysis.
As was shown in~\cite{Afonin:2012jn} and expanded upon in~\cite{Afonin:2021cwo}, the standard vector SW model~\eqref{SW}
can be generalized to the case of arbitrary intercept in the spectrum~\eqref{spSW},
\begin{equation}
\label{spSW2}
  M^2(n)=4|c|(n+1+b),
\end{equation}
where the parameter \(b\) regulates the value of intercept. The action of generalized vector SW model
takes the form (up to a multiplicative constant)
\begin{equation}
\label{gen_sw}
  S=\int d^4xdz\sqrt{g}e^{-cz^2}U^2(b,0,cz^2)\mathcal{L},
\end{equation}
where $U$ is the Tricomi hypergeometric function. The same spectrum~\eqref{spSW2} can be achieved
by introducing $z$-dependent 5D mass~\eqref{mz} with non-zero intercept parameter $b$~\cite{Afonin:2021cwo}.

Finally it should be mentioned that by a suitable choice of dilaton background $B(z)$ in the SW action~\eqref{Bz}
it is possible to reproduce any preset spectrum, even a spectrum with finite number of states~\cite{Afonin:2009xi}.
In the extensive literature on the SW holographic model, there are
numerous examples of this sort. But only the simple SW models leading to the linear Regge spectrum of the kind~\eqref{spSW2},
as predicted by stringy picture of mesons and poles of the Veneziano's amplitude,
reproduce correctly the analytic structure of OPE of correlation functions in QCD at large Euclidean momentum
(the parton logarithm plus power corrections), in this regard, these simple SW models are most related to the real QCD.
For this reason, we will consider only such SW models.

\section{Holographic Wilson loop, gluon confinement and meson string}

\begin{wrapfigure}{R}{0.4\textwidth}
  \vspace{-10mm}
  \begin{center}
    \includegraphics[width=0.38\textwidth]{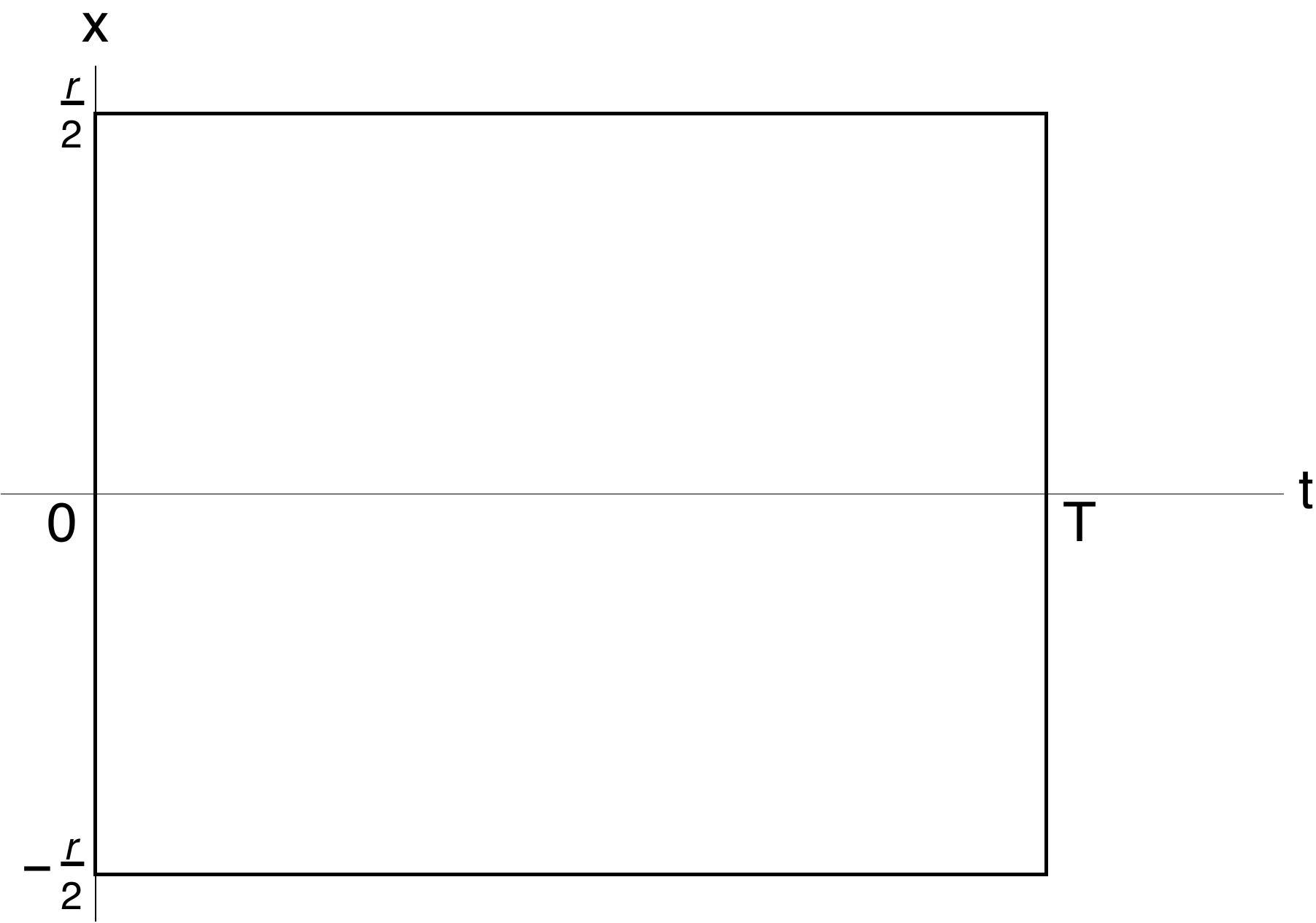}
  \end{center}
  \vspace*{-7mm}
  \caption{\footnotesize A Wilson loop.}
  \label{wilson_loop}
\end{wrapfigure}

A practical way to calculate the potential between two sources, such as~\eqref{pot}, is based on
a holographic interpretation of the expectation value of the Wilson loop suggested by
Maldacena~\cite{Maldacena:1998im}. Assume that we have a rectangular Wilson loop located in the
4D boundary of the 5D AdS space, schematically pictured in the Fig.~\ref{wilson_loop}.
Such Wilson loop can be related to the propagation of a massive quark~\cite{Maldacena:1998im}.
Then, in the limit of \(T\to\infty\), its expectation value is proportional to~\cite{Maldacena:1998im}
\begin{equation}
\label{wl_ev_1}
  \left\langle W(\mathcal{C})\right\rangle\sim e^{-TE(r)},
\end{equation}
where \(E(r)\) corresponds to the energy of the quark-antiquark pair. On the other hand,
as was proposed in~\cite{Maldacena:1998im}, it can be computed as
\begin{equation}
\label{wl_ev_2}
  \left\langle W(\mathcal{C})\right\rangle\sim e^{-S},
\end{equation}
where \(S\) represents the area of a string world-sheet which produces the loop
\(\mathcal{C}\). From comparing~\eqref{wl_ev_1} and~\eqref{wl_ev_2} one can calculate the
energy as
\begin{equation}
\label{en_def}
  E=\frac{S}{T}.
\end{equation}
A particular implementation of this idea within the SW holographic model was developed by
Andreev and Zakharov in~\cite{Andreev:2006ct}. They considered the modified Euclidean AdS\(_5\) space
with the metric~\eqref{az_metric}, in which the sign of exponent in $h$ was argued to be opposite in the
Euclidean case\footnote{Our definition of $c$ differs from the one in~\cite{Andreev:2006ct} by the factor of 4,
namely $h=e^{cz^2/2}$ in~\cite{Andreev:2006ct}. The latter leads to the absence of factor 4 in the spectrum~\eqref{spSW}.},
\begin{equation}
\label{az_metric2}
  g_{MN}=\text{diag}\left\lbrace\frac{R^2}{z^2}h,\dots,\frac{R^2}{z^2}h\right\rbrace,\quad
  h=e^{2cz^2}.
\end{equation}
The given setup was shown to result in the linearly rising energy at large distances,
\begin{equation}
\label{az_large_r_en}
  E\underset{r\to\infty}{\simeq}\sigma r + \text{Const} + \mathcal{O}(1/r),\qquad
  \sigma\equiv\frac{R^2}{\pi\alpha'}ec,
\end{equation}
where $1/\alpha'$ is the tension of fundamental string (see~\eqref{ng} below). The Coulomb behavior,
\begin{equation}
  E\underset{r\to0}{\simeq}\frac{\kappa}{r} + \text{Const} + \mathcal{O}(r),
\end{equation}
was also obtained at small distances. Thus the structure of the Cornell potential~\eqref{cornell}
was reproduced.

Our first goal is to extend this result to the case of generalized SW model with arbitrary
intercept parameter. As is seen from~\eqref{gen_sw} and~\eqref{tr}, we must apply the analysis above
to the following extension of modified Euclidean AdS metric~\eqref{az_metric2}
\begin{equation}
\label{az_metric3}
  g_{MN}=\text{diag}\left\lbrace\frac{R^2}{z^2}h,\dots,\frac{R^2}{z^2}h\right\rbrace,\quad
  h=e^{2cz^2}U^4(b,0,cz^2).
\end{equation}
In what follows we apply the analysis of~\cite{Andreev:2006ct} to the case of metric~\eqref{az_metric3}.

The starting point is the Nambu-Goto action
\begin{equation}
\label{ng}
  S=\frac{1}{2\pi\alpha'}\int d^2\xi\sqrt{\det g_{MN}\partial_\alpha X^M\partial_\beta X^N},
\end{equation}
where \(g_{MN}\) is the warped metric defined above. This action
describes the area of string world-sheet for which the expectation value of Wilson loop
should be calculated. If we choose \(\xi_1=t\) and \(\xi_2=x\) as the parametrization of the
world-sheet then, after integration over \(t\) from \(0\) to \(T\), the action
can be rewritten as
\begin{equation}
\label{ng_action}
  S=\frac{TR^2}{2\pi\alpha'}\int\displaylimits_{-r/2}^{r/2}dx\,
  \frac{h}{z^2}\sqrt{1+z'^2},
\end{equation}
where $z'=\frac{dz}{dx}$.
Since the action is translation invariant due to the absence of the explicit dependence
of the Lagrangian on \(x\), there exists a conserving quantity, which in this case is
equal to
\begin{equation}
\label{cons}
  \frac{h}{z^2}\frac{1}{\sqrt{1+z'^2}}=\text{Const}.
\end{equation}
Let us denote
\begin{equation}
  z_0\equiv\left.z\right|_{x=0},\quad
  h_0\equiv\left.h\right|_{z=z_0}.
\end{equation}
Since we know that \(z=0\) at the ends of the Wilson loop, \(x=\pm r/2\), and the system
is symmetric, $z$ reaches the maximum value of \(z=z_0\) at $x=0$. Using \(\left.z'\right|_{x=0}=0\)
we obtain from~\eqref{cons}
\begin{equation}
  \frac{h_0}{z_0^2}=\frac{h}{z^2}\frac{1}{\sqrt{1+z'^2}}\qrq
  z'=\sqrt{\frac{h^2}{h_0^2}\frac{z_0^4}{z^4}-1}\qrq
  dx=dz\frac{h_0}{h}\frac{z^2}{z_0^2}\frac{1}{\sqrt{1-\frac{h_0^2}{h^2}\frac{z^4}{z_0^4}}},
\end{equation}
and one can then write
\begin{equation}
\label{x_z_int}
  \int\displaylimits_{-r/2}^{r/2}dx=2\int\displaylimits_{0}^{z_0}dz\,\frac{h_0}{h}\frac{z^2}{z_0^2}
  \frac{1}{\sqrt{1-\frac{h_0^2}{h^2}\frac{z^4}{z_0^4}}}.
\end{equation}
The factor of 2 appears due to two equal contributions from integration from $x=-r/2$ to $0$ and from $0$ to $x=r/2$.
After defining the new variables
\begin{equation}
  v\equiv\frac{z}{z_0},\quad
  \lambda\equiv cz_0^2,
\end{equation}
and substituting \(h(z)\) we get the final expression for the distance $r$,
\begin{equation}
  r=2\sqrt{\frac{\lambda}{c}}\int\limits_0^1dv\,
  \frac{U^4(b,0,\lambda)}{U^4(b,0,\lambda v^2)}
  \frac{v^2e^{2\lambda(1-v^2)}}
  {\sqrt{1-v^4e^{4\lambda(1-v^2)}\frac{U^8(b,0,\lambda)}{U^8(b,0,\lambda v^2)}}}.
\end{equation}

The expression for the energy can be obtained from the equation~\eqref{en_def} and the
action~\eqref{ng_action} by replacing integration over \(x\) with the integration over
\(z\) using~\eqref{x_z_int}. The resulting expression, however, is divergent at \(v=z=0\)
and thus requires regularization. After imposing the cutoff $\veps\rightarrow0$ we get the
regularized energy
\begin{equation}
  E_R=\frac{R^2}{\pi\alpha'\veps}+E,
\end{equation}
with the energy integral equal to
\begin{equation}
  E=\frac{R^2}{\pi\alpha'}\sqrt{\frac{c}{\lambda}}\lsb\int\limits_0^1\frac{dv}{v^2}\lb
  \frac{e^{2\lambda v^2}U^4(b,0,\lambda v^2)}{\sqrt{1-v^4e^{4\lambda(1-v^2)}\frac{U^8(b,0,\lambda)}{U^8(b,0,\lambda v^2)}}}-D\rb-D\rsb,
\end{equation}
where we introduced the regularization constant \(D\equiv U^4(b,0,0)\). At this point we have
expressed the potential in terms of the parametric functions \(r(\lambda)\) and
\(E(\lambda)\). In order to find the asymptotics \(r\to\infty\) we need to express it in terms of \(\lambda\).

\begin{wrapfigure}{r}{0.4\textwidth}
  \vspace{-5mm}
  \begin{center}
    \includegraphics[width=0.38\textwidth]{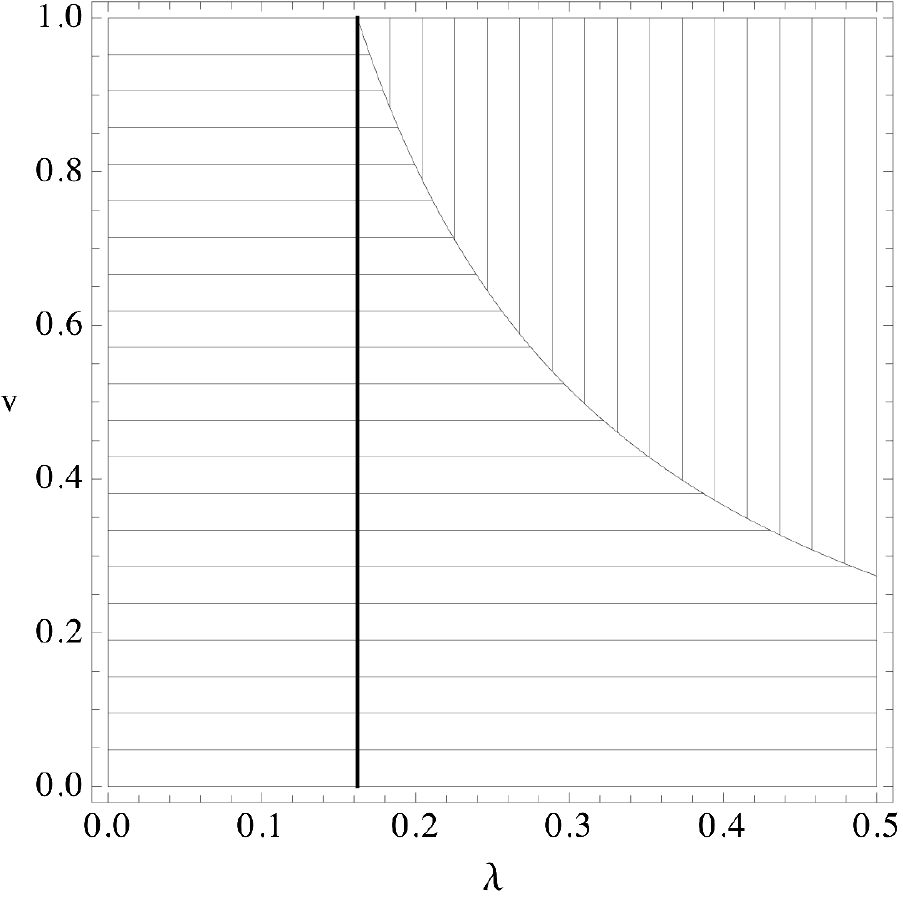}
  \end{center}
  \vspace*{-7mm}
  \caption{\footnotesize The contour plots of the expression under the square root inside
  the integrals, see~\eqref{sqrt_exp}, in the case of \(b\approx-0.48\). The black vertical
 line is \(\lambda\approx 0.16\). The horizontal lines display the region where~\eqref{sqrt_exp}
is non-negative. The region of negative~\eqref{sqrt_exp} is marked by vertical lines.}
  \label{real_cond_b05}
\end{wrapfigure}

Let us start by noting that these integrals must be real-valued which requires the
expression under the square root to be non-negative in the entire integration range
\(v\in\lsb 0,1\rsb\),
\begin{equation}
\label{sqrt_exp}
  1-v^4e^{4\lambda(1-v^2)}\frac{U^8(b,0,\lambda)}{U^8(b,0,\lambda v^2)}\ge0.
\end{equation}
In its turn, this condition is satisfied if the minimum value of this expression is non-negative
within the same range. The minima are given by the roots of the derivative of~\eqref{sqrt_exp}.
For $v\neq0$, this leads to the equation
\begin{equation}
\label{eq12}
        1-2\lambda v^2-4\lambda v^2\dfrac{U'(b,0,\lambda v^2)}{U(b,0,\lambda v^2)}=0, \qquad
    U(b,0,\lambda v^2)\ne 0,
\end{equation}
where $U'(b,0,x)=\partial_x U(b,0,x)$.
Denoting
\begin{equation}
  x\equiv\lambda v^2,
\end{equation}
the numerical calculations show that the integrals are real-valued if
\begin{equation}
  \lambda<x.
\end{equation}
In Fig.~\ref{real_cond_b05}, we display an example of behavior of the expression~\eqref{sqrt_exp}
as the function of two arguments, \(\lambda\) and \(v\). The expression is non-negative in the region to the
left of the thick vertical line, hence, the
integrals are real-valued for all \(v\in\lsb0,1\rsb\) in that region.

It is interesting to remark that similar conditions can be obtained from the Sonnenschein confinement
criterion~\cite{Sonnenschein:2000qm} which imposes the following requirements on the \(g_{00}\) element of the
warped metric
\begin{equation}
\label{sonn_cond}
  \left.\pz g_{00}\right|_{z=z_0}=0,\quad
  \left.g_{00}\right|_{z=z_0}\ne0,
\end{equation}
in order for the background to be dual to a confining theory in the sense of area law behavior of
Wilson loop. In our case, using the
previous notation \(\lambda=cz_0^2\) and after certain simplifications, the second of
Sonnenschein conditions takes the form
\begin{equation}
\label{cond1}
  U(b,0,\lambda)\ne 0.
\end{equation}
The first condition reads as follows
\begin{equation}
  \frac{U^4(b,0,\lambda)}{(\lambda/c)^{3/2}}\lb1-2\lambda-4\lambda\frac{U'(b,0,\lambda)}{U(b,0,\lambda)}\rb=0.
\end{equation}
Due to the first condition the only possible case when this equality can be true is
\begin{equation}
\label{cond2}
  1-2\lambda-4\lambda\frac{U'(b,0,\lambda)}{U(b,0,\lambda)}=0.
\end{equation}
The conditions~\eqref{cond1} and~\eqref{cond2} coincide with the equation~\eqref{eq12}.

Going back to the integrals we note that since \(r\) is a growing function of \(\lambda\),
for obtaining the large distance asymptotics we can consider the ``large'' \(\lambda\) asymptotics.
Since \(\lambda\) is bounded from above, this means finding the asymptotics \(\lambda\to x\) .

Further we follow the procedure outlined in~\cite{Andreev:2006ct}. Since
the integrals diverge in the upper bound \(v=1\), we expand the expression under
the square root near \(v=1\). The integrals transform into
\begin{equation}
\label{rbeh}
  r\underset{v\to1}=2\sqrt{\frac{\lambda}{c}}\int\limits_0^1\frac{dv}{\sqrt{A(b,\lambda)(1-v)+B(b,\lambda)(v-1)^2}},
\end{equation}
\begin{equation}
\label{Ebeh}
  E\underset{v\to1}=\frac{R^2}{\pi\alpha'}\sqrt{\frac{c}{\lambda}}e^{2\lambda}U^4(b,0,\lambda)
  \int\limits_0^1\frac{dv}{\sqrt{A(b,\lambda)(1-v)+B(b,\lambda)(v-1)^2}},
\end{equation}
where \(A\) and \(B\) are some functions of \(b\) and \(\lambda\) which do
not depend on \(v\) (see Appendix~A).  We do not need the analytical expressions for $r$ and $E$
(see Appendix~A) as we are interested in the behavior $E(r)$.
Substituting~\eqref{rbeh} into~\eqref{Ebeh} we get
\begin{equation}
\label{large_r_en}
  E\underset{r\to\infty}{\sim}\frac{R^2}{2\pi\alpha'}\frac{e^{2x}U^4(b,0,x)}{x}cr,
\end{equation}
where  \(\lambda\to x\) while \(x\) is the solution of the equation
\begin{equation}
\label{x_eq}
  1-2x-4x\dfrac{U'(b,0,x)}{U(b,0,x)}=0,
\end{equation}
which can be found numerically.

As follows from~\eqref{pot}, the coefficient in front of \(r\) in~\eqref{large_r_en}
represents the string tension,
\begin{equation}
\label{gen_vector_sigma}
  \sigma(b)=\frac{R^2}{2\pi\alpha'}\frac{e^{2x}U^4(b,0,x)}{x}c,
\end{equation}
where the dependence on the parameter \(b\) is hidden inside \(x\) by virtue of the
equation~\eqref{x_eq}. Since \(U(0,0,x)=1\), the equation~\eqref{x_eq} yields \(x=1/2\) at $b=0$
and thus~\eqref{gen_vector_sigma} reproduces the string tension~\eqref{az_large_r_en} of
Andreev-Zakharov analysis in~\cite{Andreev:2006ct},
\begin{equation}
\label{vector_sigma}
  \sigma(0)=\frac{R^2}{\pi\alpha'}ec.
\end{equation}

\begin{wrapfigure}{R}{0.4\textwidth}
  \vspace{-7mm}
  \begin{center}
    \includegraphics[width=0.38\textwidth]{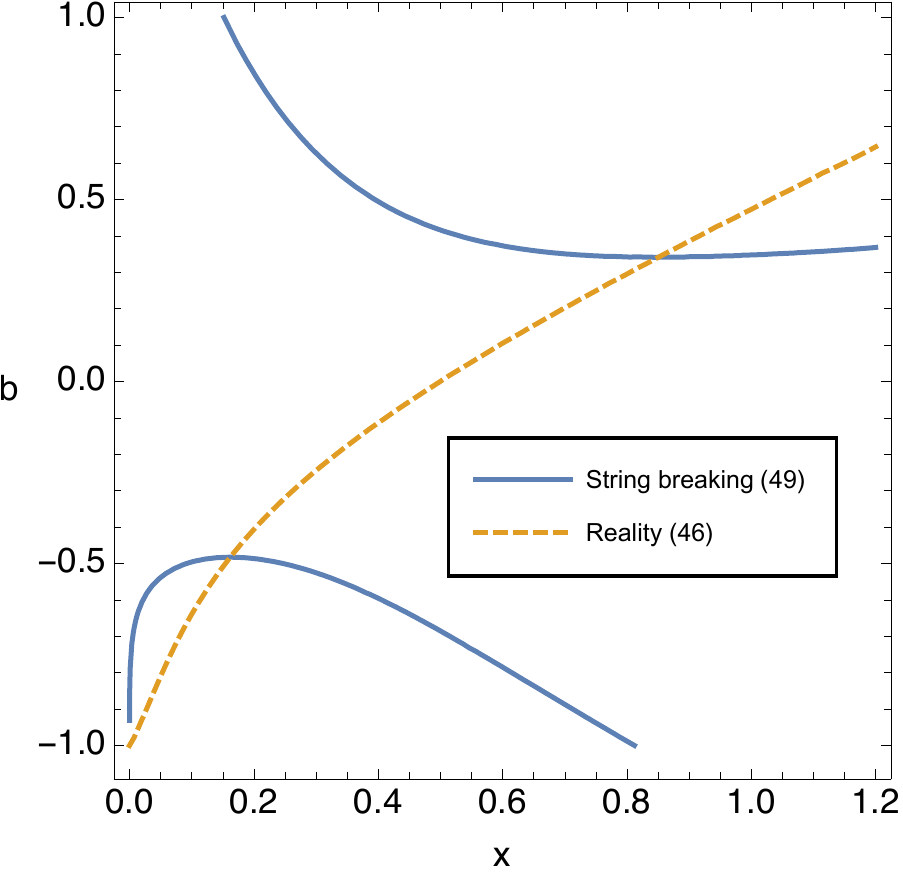}
  \end{center}
  \vspace*{-7mm}
  \caption{\footnotesize The string breaking (blue solid) and reality (orange
  dashed) conditions in the vector case.}
  \vspace*{1mm}
  \label{sys_sols_vector}
\end{wrapfigure}

We approached to the key point of our analysis. Following the discussion in the Section~1, we
interpret the case of $b=0$ as corresponding to pure gluodynamics that absolutely dominates
in the large-$N_c$ limit --- the limit where the holographic models are supposedly defined.
Thus the case of $b=0$ (zero ``quark'' external parameter) is interpreted as the case of closed gluon
string with pure gluon sources. Now we impose the closed string breaking condition~\eqref{sigma_ratio}.
From~\eqref{gen_vector_sigma} and~\eqref{vector_sigma} we get the equation
\begin{equation}
\label{strbr}
  \frac{e^{2x-1}U^4(b,0,x)}{2x}=\frac{1}{2}.
\end{equation}

The equations~\eqref{x_eq} (the reality condition) and~\eqref{strbr} (the closed string breaking condition) form a system of two equations for two parameters
\(b\) and \(x\). As we consider the vector case, the solutions $b_0$ for the intercept parameter are interpreted as our prediction
for the vector spectrum of quark-antiquark pairs.

The closed string breaking condition~\eqref{strbr} has two branches and each one is intersected by the reality condition~\eqref{x_eq}, see Fig.~\ref{sys_sols_vector}
(according to the Eq.~\eqref{strbr}, the seeming intersection at $x=0$ is not a solution\footnote{Actually the asymptotically emerging point $(x,b)=(0,-1)$
seems to require a separate study. The arising spectrum would correspond to the deep ultraviolet limit $z\rightarrow0$ (as $x=cz^2$) and, according to~\eqref{spectrum2},
would contain a massless vector particle. A physical interpretation (the photon?) is unclear.}).

The first numerical solution (intersection with the lower branch) yields
\begin{equation}
\label{sol1}
  b_0\approx-0.483,\qquad
  x\approx0.162.
\end{equation}
The obtained value of $b_0$ in~\eqref{sol1} is remarkably close to \(b=-0.5\) which arises
for vector mesons in various approaches. Namely, according to~\eqref{spectrum2}
the predicted radial spectrum of \(\rho\)-mesons (and of their isoscalar counterparts $\omega$)
is close to
\begin{equation}
\label{rho_spectrum}
  M_\rho^2(n)=2M_\rho^2\lb n+\frac{1}{2}\rb,
\end{equation}
where \(M_\rho\) is the mass of the ground \(\rho\)-meson state $\rho(770)$~\cite{pdg}.
The spectrum~\eqref{rho_spectrum} emerged for the first time in the Veneziano dual amplitudes
after imposing the Adler self-consistency condition~\cite{Ademollo:1969nx,Collins:1971ff,Collins:1977jy}
(the pion scattering amplitude is zero at zero momentum).
The relation~\eqref{rho_spectrum} also appeared
in the finite energy QCD sum rules~\cite{Krasnikov:1981vw} and in the QCD sum rules
in the large-$N_c$ limit~\cite{Afonin:2003gp,Afonin:2004yb,Afonin:2006da}.
This relation arises naturally in the light-front holographic approach~\cite{br3}.
The experimental data on the form-factor of charged pion $F_\pi(Q^2)$
have the most satisfactory resonance description if the radial vector spectrum is close to the behavior
$m_\rho^2(n)\sim n+1/2$ as in~\eqref{rho_spectrum} (see, e.g.,~\cite{Afonin:2021cwo,br3}).
Within the framework
of essentially the same holographic model, a very close value of $b_0$ was recently obtained from
fitting the mean pion charge radius squared~\cite{Afonin:2021cwo}. We see thus that the proposed new
approach passes an important phenomenological test.

Taking the standard mean value for the slope
of meson radial trajectories $a'=4|c|\approx1.14$~GeV$^2$~\cite{Bugg:2004xu} in~\eqref{spectrum2}, the
predicted large-$N_c$ masses are $M(n)\approx\{0.77,1.32,1.69,2.00,2.27,\dots\}$~GeV. It looks natural to match
them to the resonances $\rho(770)$, $\rho(1450)$ (it is a broad resonance region including possible $\rho(1270)$~\cite{pdg}),
$\rho(1700)$, $\rho(2000)$, and $\rho(2270)$, correspondingly~\cite{pdg}.

The second numerical solution (intersection with the upper branch) gives
\begin{equation}
\label{sol2}
  b_0\approx0.341,\qquad
  x\approx0.851.
\end{equation}
The same input for the slope leads to the radial spectrum
$M(n)\approx\{1.24,1.63,1.95,2.22,\dots\}$~GeV. This spectrum is
surprisingly close to the experimental spectrum of axial-vector mesons $a_1$
--- the chiral partners of $\rho$-mesons:  $a_1(1260)$ (the mass is $1230\pm40$~GeV),
$a_1(1640)$, $a_1(1930)$, and $a_1(2270)$, correspondingly~\cite{pdg}.
The degree of agreement with experiment is quantified in the Appendix~B, where we performed
an RMS analysis which shows that the corresponding RMS error is small (at the level of several percents).

We propose that the observed coincidence is not accidental but points at a new way of
incorporation of the Chiral Symmetry Breaking (CSB) into the bottom-up holographic approach.

We remind the reader that the AdS$_5$ space locally looks like the 5D Minkowski space, i.e.,
a space with one time and four flat space directions. The space rotations in four dimensions
contain two independent three-dimensional rotations, $SO(4)=SO(3)\times SO(3)$. Correspondingly,
the spin is marked by two numbers. In the vector case, this leads to the existence of two spin-1
mesons --- the ``Left'' one $A^M_L$ with spin $(1,0)$ and the ``Right'' one $A^M_R$ with
spin\footnote{There is also the representation $(1/2,1/2)$ but such a state does not carry helicity $\lambda$
(according to a Weinberg theorem, the Lorentz representation $(a,b)$ carries the helicity $\lambda=|b-a|$)
describing just a derivative of a scalar field with only one physical degree of freedom.} $(0,1)$.
From the viewpoint of representations of Lorentz group,
$SO(3)$ is the rotational subgroup of the little group for 5D massless particles (the extension of $SO(2)$
for 4D massless particles) which coincides with the little group for 4D massive particles. The ensuing equality
of physical degrees of freedom makes possible a direct Kaluza-Klein like projection of 5D massless particles
to 4D massive ones. But the 5D states have no definite space parity (this notion appears only for an odd number of
space dimensions), hence, the dimensional reduction must be performed together with constructing states of definite
space parity. In the spin-1 case under consideration, the corresponding states are the usual vector,
$V^\mu=(A^\mu_R+A^\mu_L)/\sqrt{2}$, and the axial-vector, $A^\mu=(A^\mu_R-A^\mu_L)/\sqrt{2}$, as now the
$P$-reflection interchanges $(1,0)$ and $(0,1)$. The Lagrangian density for free vector fields in~\eqref{SW}
becomes
\begin{equation}
\mathcal{L}=-\frac14F^2=-\frac14\left(F^2_L+F^2_R\right)=-\frac14\left(F^2_V+F^2_A\right).
\end{equation}
The field $V_\mu$ for quark-antiquark states is supposedly dual to the QCD operator $\bar{q}\gamma_\mu q$
or its isovector partner $\bar{q}\gamma_\mu\vec{\tau} q$ and $A_\mu$ is dual to $\bar{q}\gamma_\mu\gamma_5 q$
or $\bar{q}\gamma_\mu\gamma_5\vec{\tau} q$.
Since these operators have equal mass dimension $\Delta=3$, for description of mass splitting between parity
partners due to the CSB effects one needs some {\it ad hoc} modification that usually has no motivation inside
the bottom-up holographic approach itself. The most known examples are incorporation of pions in the spirit of chiral
perturbation theory in the pioneering Hard-Wall models~\cite{son1,pom} and incorporation of quark angular momentum
in the light-front holographic QCD~\cite{br3}.

We conjecture that the existence of two branches in the closed string breaking condition~\eqref{strbr}
corresponds to the possibility of having two kinds of spin-1 states, the vector and axial mesons, after
the dimensional reduction of 5D fields $A^M_L$ and $A^M_R$. We obtain then a description of mass splitting between
the parity partners without any {\it ad hoc} modifications
based on a ``pre-understanding'' of CSB from other approaches.
Remarkably, the arising description turns out to be quantitative.

\section{Scalar case}

The original Andreev-Zakharov analysis~\cite{Andreev:2006ct} was based on the vector SW model.
It is interesting to apply this analysis and our extension to the scalar case (the SW model
describing the scalar fields). According to the general design of generalized SW models with
arbitrary intercept, the Tricomi function in~\eqref{gen_sw} is replaced by $U(b,-1,cz^2)$ in
the scalar case~\cite{Afonin:2021cwo} and the SW scalar spectrum is (see, e.g.,~\cite{Afonin:2021cwo})
\begin{equation}
\label{spSW3}
  M^2(n)=4|c|(n+3/2+b).
\end{equation}
The transformation~\eqref{tr} then leads to the following
background function in the metric~\eqref{az_metric3},
\begin{equation}
  h=e^{2cz^2/3}U^{4/3}(b,-1,cz^2).
\end{equation}
Repeating the same steps as in the vector case, we obtain the following expressions for
the distance \(r(\lambda)\) and energy \(E(\lambda)\)
\begin{equation}
  r=2\sqrt{\frac{\lambda}{c}}\int\limits_0^1dv\,
  \frac{U^{4/3}(b,-1,\lambda)}{U^{4/3}(b,-1,\lambda v^2)}
  \frac{v^2e^{2\lambda(1-v^2)/3}}
  {\sqrt{1-v^4e^{4\lambda(1-v^2)/3}\frac{U^{8/3}(b,-1,\lambda)}{U^{8/3}(b,-1,\lambda v^2)}}},
\end{equation}
\begin{equation}
  E=\frac{R^2}{\pi\alpha'}\sqrt{\frac{c}{\lambda}}\lsb\int\limits_0^1\frac{dv}{v^2}\lb
  \frac{e^{2\lambda v^2/3}U^{4/3}(b,-1,\lambda v^2)}{\sqrt{1-v^4e^{4\lambda(1-v^2)/3}\frac{U^{8/3}(b,-1,\lambda)}{U^{8/3}(b,-1,\lambda v^2)}}}-D\rb-D\rsb,
\end{equation}
with the regularization constant \(D\equiv U^{4/3}(b,-1,0)\). They
lead to the large distance asymptotics for the energy
\begin{equation}
  E\underset{r\to\infty}{\sim}\frac{R^2}{2\pi\alpha'}\frac{e^{2x/3}U^{4/3}(b,-1,x)}{x}cr.
\end{equation}
The parameter \(x\) has the same meaning as in the vector case --- it is
one of the roots of derivative of the expression under the square root in the integrals above.
The scalar counterpart of the reality condition~\eqref{x_eq} becomes
\begin{equation}
\label{realscalar}
  1-\frac{2}{3}x-\frac{4}{3}x\frac{U'(b,-1,x)}{U(b,-1,x)}=0,
\end{equation}
and the one for the string tension~\eqref{gen_vector_sigma} is
\begin{equation}
\label{stringscalar}
  \sigma(b)\equiv\frac{R^2}{2\pi\alpha'}\frac{e^{2x/3}U^{4/3}(b,-1,x)}{x}c.
\end{equation}
The particular case $b=0$ corresponds to
\begin{equation}
\label{scalar_sigma}
  \sigma(0)=\frac{R^2}{3\pi\alpha'}ec.
\end{equation}
It differs from~\eqref{vector_sigma} by the factor of $1/3$. This difference arises from the
formulation of SW model in the form of~\eqref{az_metric2}: If, for the purpose of describing
the Regge form of spectrum, one introduces an IR modification in the metric, the slope becomes
spin-dependent (actually this was the reason to introduce the IR modification in the form of
dilaton background $e^{-cz^2}$ in the original paper by Son et al.~\cite{son2}). According to
the transformation~\eqref{tr}, the function $e^{cz^2}$ in the Euclidean dilaton background
becomes $e^{2cz^2}$ in the spin-1 case and $e^{2cz^2/3}$ in the spin-0 one resulting in the
factor of $1/3$ that emerged in~\eqref{scalar_sigma}.

\begin{wrapfigure}{R}{0.4\textwidth}
  \vspace{-7mm}
  \begin{center}
    \includegraphics[width=0.38\textwidth]{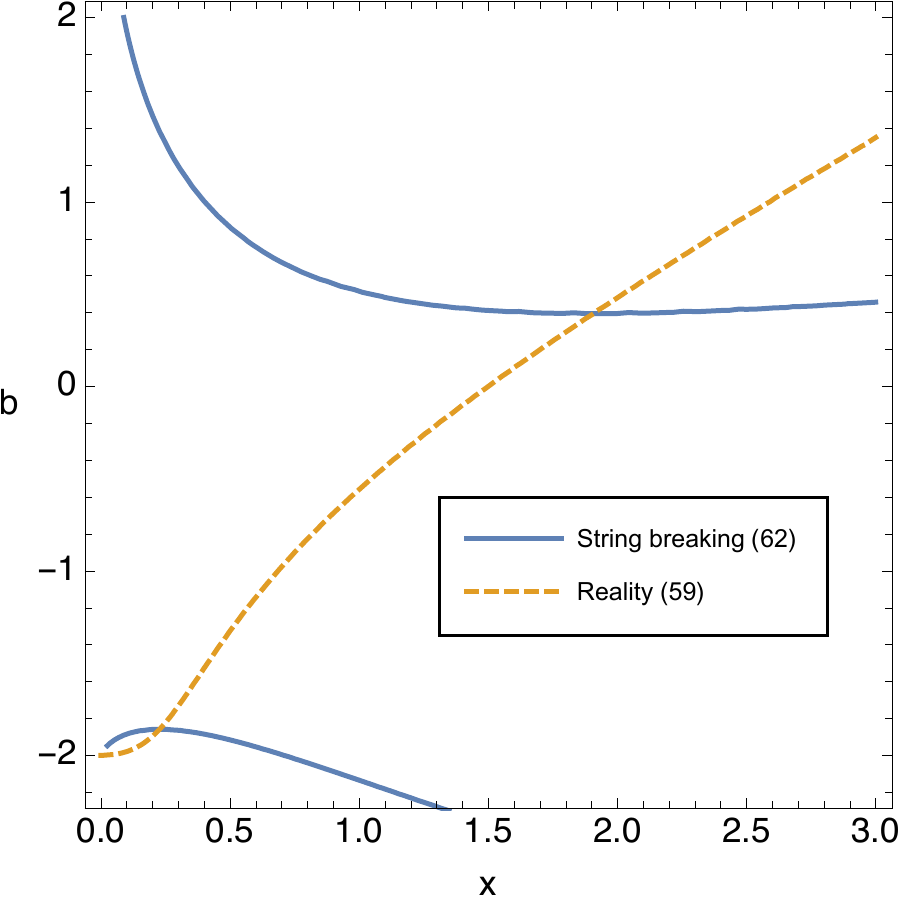}
  \end{center}
  \vspace*{-7mm}
  \caption{\footnotesize The string breaking (blue solid) and reality (orange
  dashed) conditions in the scalar case.}
  \vspace*{2mm}
  \label{sys_sols_scalar}
\end{wrapfigure}

As follows from~\eqref{stringscalar} and~\eqref{scalar_sigma}, our condition for the closed gluon
string breaking~\eqref{sigma_ratio} (which does not depend on a
general constant factor in $\sigma$) now reads
\begin{equation}
\label{scalar_breaking}
      \dfrac{3e^{2x/3-1}U^{4/3}(b,-1,x)}{2x}=\dfrac{1}{2}.
\end{equation}
As in the vector case, this condition has two branches, see Fig.~\ref{sys_sols_scalar},
and we may try to interpret them as arising for scalar states of opposite
space parity, namely the lower branch should correspond to mesons
of negative parity and the upper one to heavier mesons of positive parity.

The reality condition~\eqref{realscalar} intersects the upper branch in the point
\begin{equation}
\label{scalar_intercept1}
  b_0\approx0.394,\qquad x\approx1.906.
\end{equation}
The scalar sector is notoriously difficult for theoretical descriptions, especially
the scalar isoscalar resonances $f_0$ having the vacuum quantum numbers. Many of
them are believed to be not dominated by the quark-antiquark component~\cite{pdg}.
If true, this means that a pattern of spectrum emerging in the large-$N_c$ limit
can be very different from the reality. Since in our holographic approach this limit is supposed to be built-in,
the comparison with the isovector scalar
mesons $a_0$ looks more reasonable but the corresponding experimental data are scarce.
Nevertheless, the spectrum~\eqref{spSW3} with the standard mean slope $4|c|=1.14$~GeV$^2$~\cite{Bugg:2004xu}
and the intercept $b_0$ given by the solution~\eqref{scalar_intercept1} predicts for the ground state the mass
$M(0)\approx1.47$~GeV which matches perfectly the mass of $a_0(1450)$-meson (its experimental mass is
$1474\pm19$~GeV~\cite{pdg}). This agrees with a widespread opinion
about tetraquark nature of $a_0(980)$ (together with its isoscalar counterpart
$f_0(980)$) and that the lowest quark-antiquark $a_0$-meson is $a_0(1450)$~\cite{pdg}.

The lower branch in Fig.~\ref{sys_sols_scalar} is intersected by the reality condition~\eqref{realscalar} in the point
\begin{equation}
\label{scalar_intercept2}
  b_0\approx-1.859,\qquad x\approx 0.227.
\end{equation}
The ground state of spectrum~\eqref{spSW3} is then tachyonic. This looks like a failure of our approach.
On the other hand, the scalar SW spectrum~\eqref{spSW3} was derived for the fields supposedly dual to the
scalar QCD operators of canonical dimension $\Delta=3$, which are $\bar{q}q$ and $\bar{q}\gamma_5q$. If we
do not specify $\Delta$, the SW scalar spectrum is~\cite{afonin2020}
\begin{equation}
\label{spSW4}
  M^2(n)=4|c|(n+\Delta/2+b).
\end{equation}
As is known from the old Gell-Mann--Oakes--Renner relation, the pion mass is related with the renormalization-group invariant QCD operator
$m_q\bar{q}q$ that has the mass dimension $\Delta=4$. The spectrum~\eqref{spSW4} yields then
$M(0)\approx0.4$~GeV which already somewhat resembles the physical pion mass\footnote{As in the vector case, the apparent intersection at $x=0$
in Fig.~\ref{sys_sols_scalar} is not a solution to the string breaking condition~\eqref{scalar_breaking}. But one can approach infinitely close to
the point $(x,b)=(0,-2)$ which for $\Delta=4$ in~\eqref{spSW4} would lead to a massless pseudoscalar particle in the deep ultraviolet limit.
A physical interpretation (the pion?) is again unclear.}.
The estimated value of mass $0.4$~GeV, however, seems to suggest a more natural interpretation for this light scalar state ---
a very close mass has the two-pion scalar resonance called sigma-meson or $f_0(500)$ in Particle Data~\cite{pdg}.

\section{Discussions}

Our analysis above was based on connecting the holographic Regge trajectories for gluons and mesons, expressions~\eqref{spectrum1} and~\eqref{spectrum2}, making use of gluon dominance in strong dynamics as prescribed by the large $N_c$ (planar) limit in QCD~\cite{hoof,wit}. A natural question that may arise with regard to our analysis is why a heavy meson potential, like the one coming from Wilson loops, was used for light mesons? In the original considerations by Maldacena~\cite{Maldacena:1998im}, the Chan-Paton factors at the boundary were considered at rest and taken as the heavy quark rest mass, implying quark kinematics to be neglected. For light mesons, however, the relativistic kinetic energy part should be considered.
This question seems to be closely related with the old question why the non-relativistic potential models work reasonably not only for heavy-quark systems
but also, usually after some relativistic modifications, for light-quark systems? A completely satisfactory answer to this question is unknown, the only reliable justification of the
application of potential models to bound states of light quarks are their phenomenologically successful predictions. There exist, however, various qualitative
insights showing that the ability of potential models to give semiquantitative results for light-quark spectroscopy is certainly not accidental, but can be traced back
to some physical arguments, the corresponding discussions are presented in detail in the review~\cite{Lucha}. The general lesson is that in calculating
the bound energies using some potential, like the Cornell potential, a universal non-perturbative mass scale should be used both in the heavy and light quark sector.
This scale manifests itself as universal slope $\sigma$ of linearly rising potential. The light quarks acquire a dynamical constituent mass $M_\text{dyn}\sim\sqrt{\sigma}\sim350$~MeV
effectively becoming sufficiently heavy degrees of freedom to be treated non-relativistically.

There are more modern phenomenological hints from the connection of Regge trajectories in the heavy and light sector.
As was first demonstrated in Ref.~\cite{Afonin:2014nya}, the extension of radial Regge spectrum~\eqref{spectrum2}
to the case of vector mesons containing heavy quarks of mass $m_1$ and $m_2$ is very well approximated by
\begin{equation}
\label{spectrum2b}
  \left(M_\text{mes}(n)-m_1-m_2\right)^2=a'(n+1+b),
\end{equation}
with essentially the same slope $a'$ and even intercept $b$ as for the light mesons. This observation suggests that
if the quark bound energies are calculated from some eigenvalue problem, e.g., as eigenvalues from a local potential,
the masses of light mesons look like almost pure bound energies, while the heavy quarks add a contribution to masses,
in the first approximation, in a simple way displayed in~\eqref{spectrum2b}.

What is most important in these examples is that the idea of finding the parameters of light meson spectrum in the heavy
quark limit is not unreasonable. Actually essentially this assumption was implicit in the original analysis of the holographic
Wilson loop by Andreev and Zakharov~\cite{Andreev:2006ct} which we used in the present work.

A substantial point in our analysis was the matching of two different realizations of confinement. In the standard SW model,
confinement is achieved by the existence of bounded states dual to hadrons organized in linear Regge trajectories.
In the stringy consideration, the Wilsonian loop becomes a kind of confinement parameter order. In both cases, one can construct
a dual holographic description. In the first case, the meson Regge trajectory comes from the eigenvalues of a holographic potential
that is constructed from an object dual to two-point functions, while in the second one, the Cornell-like potential comes from an object
dual to a Wilson loop. {\it A priori} it is not evident why the slope at the holographic Cornell potential should be equivalent to the
holographic Regge slope in the SW model because it is not evident that two different dual objects should be equivalent at the bulk.
But as in the aforementioned case of potential description of heavy and light quark systems, we assume that both considerations
share the same non-perturbative mass scale which controls the scale of meson masses. Our matching, in a sense, consisted in identifying
the mass scale present in both dual descriptions of confinement physics.

\section{Conclusion}

A vital problem for any successful phenomenological model is fixing its parameters from some
theoretical principles. We considered a generalized version of phenomenologically successful
holographic Soft Wall model~\cite{son2} that contains an extra ``intercept'' parameter $b$ contributing
both to the Regge like mass spectrum and to other observables like hadron form factors~\cite{Afonin:2021cwo},
and proposed a theoretical way for calculation of this parameter within the framework of holographic
approach itself. Our method is based on a holographic realization of the Wilson confinement criterion
and the introduced closed string breaking condition.
We applied this method to the vector and scalar cases and found a very good agreement with many other
phenomenological approaches and with the experimental data on light unflavored mesons. We argued that
within the framework of the proposed holographic approach, the effects of chiral symmetry breaking seem
to be built-in, at least the mass splitting between the parity partners may be described without
incorporation of additional modifications.

The proposed model can be extended to the case of arbitrary integer spin.
It would be also interesting to apply our approach,
with certain, perhaps serious, modifications, to the actively developing holographic Higgs sector ---
it is not excluded that the condition for the creation of quark-antiquark states that we used might have
a counterpart for the creation of fermion-antifermion pairs from the Higgs field and this could be exploited
for a holographic prediction, e.g., of the Higgs mass.

\section*{Acknowledgements}
This research was funded by the Russian Science Foundation grant
number 21-12-00020.

\section*{Appendix A}

Introducing the notation
\begin{equation}\label{s_func}
  s(b,\lambda;v)\equiv 1-v^4e^{4\lambda(1-v^2)}\frac{U^8(b,0,\lambda)}{U^8(b,0,\lambda v^2)},
\end{equation}
the functions \(A\) and \(B\) in the expressions~\eqref{rbeh} and~\eqref{Ebeh} can be written as
\begin{equation}
  A(b,\lambda) = -s'_v(b,\lambda;1)=
  4\lsb 1-2\lambda-4\lambda\frac{U'(b,0,\lambda)}{U(b,0,\lambda)}\rsb,
\end{equation}
\begin{multline}
  B(b,\lambda) = \frac{s''_{vv}(b,\lambda;1)}{2}=-2\lsb 16\lambda^2-18\lambda+3+
  72\lambda^2\frac{U'(b,0,\lambda)^2}{U(b,0,\lambda)^2}\,+\right.\\\left.+\,
  4\lambda\frac{(16\lambda-9) U'(b,0,\lambda)-2 \lambda U''(b,0,\lambda)}{U(b,0,\lambda)}\rsb.
\end{multline}
The integral in the expressions~\eqref{rbeh} and~\eqref{Ebeh} can be calculated analytically using~\cite{GradRyzh}
\begin{equation}
  \int\frac{du}{\sqrt{Au+Bu^2}}=\frac{1}{\sqrt{B}}\ln\lb2\sqrt{B(Au+Bu^2)}+2Bu+A\rb.
\end{equation}
This integral produces logarithmic in \(\lambda\) asymptotics.

\section*{Appendix B}

We compute RMS for Vector (V) and Axial (A) spectra of light isovector mesons using the standard definition,
\begin{equation}
    \varepsilon_\text{RMS}=\sqrt{\frac{1}{n_\text{obs}-n_\text{par}}\sum_{i=1}^{n_\text{obs}}\lb\frac{m_{\text{exp},i}-m_{\text{th},i}}{m_{\text{exp},i}}\rb^2},
\end{equation}
where \(n_\text{obs}\) is the number of observables (masses), \(n_\text{par}\) is the number of
parameters, \(m_\text{exp}\) represents the experimental value of a mass, and \(m_\text{th}\) denotes
its theoretical prediction. The meson spectrum under consideration has the linear Regge form
\begin{equation}
  m_n^2=a(n+1+b),
\end{equation}
where the phenomenological value for the slope is taken equal to \(a=1.14\,\text{GeV}^2\).
The values of \(m_\text{exp}\) from PDG~\cite{pdg} and \(m_\text{th}\) are collected in the table below
\[\begin{array}{c|c|c}
  & m_\text{exp}, \text{GeV} & m_\text{th}, \text{GeV} \\
  \hline
  \rho(770)  & 0.78 & 0.77 \\
  \rho(1450) & 1.47\pm 0.03 & 1.32 \\
  \rho(1700) & 1.72\pm 0.02 & 1.69 \\
  \rho(2000) & 2.00\pm 0.03 & 2.00 \\
  \rho(2270) & 2.27\pm 0.04 & 2.27 \\
  \hline\hline
  a_1(1260) & 1.23\pm 0.04 & 1.24 \\
  a_1(1640) & 1.66\pm 0.02 & 1.63 \\
  a_1(1930) & 1.93^{+0.03}_{-0.07} & 1.95 \\
  a_1(2270) & 2.27^{+0.06}_{-0.04} & 2.22 \\
\end{array}\]
The corresponding RMS errors for deviation from the ``central'' experimental values are
\begin{equation}
  \varepsilon^V_\text{RMS}=6.0\%,\quad
  \varepsilon^{A}_\text{RMS}=2.2\%.
\end{equation}
Since the resonance masses are usually spread over some interval
(see ~\(\rho(1450)\), \(\rho(1700)\), \(a_1(1640)\), and \(a_1(2270)\) in the table above), it makes sense to present also
the RMS for deviation from the ends of the corresponding intervals, the result is
\begin{equation}
  \varepsilon^V_\text{RMS}=4.9\%,\quad
  \varepsilon^{A}_\text{RMS}=1.1\%.
\end{equation}
Taking into consideration the typical accuracy of the large-$N_c$ limit in QCD, of the order of 10--20\%,
the accuracy of predictions above look surprisingly good.

\end{document}